\documentclass[11pt]{article}
\usepackage{amssymb,amsfonts,amsmath, amsthm, amsbsy, rotating}
\usepackage{caption,color,graphicx,paralist, subcaption, placeins, array, mathtools, url, mdwlist, color, tikz ,lineno} 
\usepackage[text={6.75in,9.5in},centering,letterpaper]{geometry}
\usepackage[linktoc=all,hypertexnames=false,colorlinks=true,urlcolor=blue,linkcolor=blue,citecolor=blue]{hyperref}
\usepackage[shortlabels]{enumitem}
\usepackage{csquotes}
\usetikzlibrary{decorations.pathmorphing, calc, arrows.meta}

\usepackage{setspace}
\usepackage[round, semicolon]{natbib}
\usepackage{booktabs}
\usepackage{authblk}
\usepackage[labelfont=bf]{caption}

\everymath={\displaystyle}
 \numberwithin{equation}{section}
 
\usepackage[thinc]{esdiff}

\title{Learning Climate Sensitivity from Future Observations, Fast and Slow}
\author[1,2]{Adam Michael Bauer\thanks{Corresponding author email: ambauer@uchicago.edu}}
\author[3,4]{Cristian Proistosescu}
\author[3,5]{Kelvin K Droegemeier}
\affil[1]{\small \textit{Department of Physics, University of Illinois Urbana-Champaign, Loomis Laboratory, Urbana, Illinois}}
\affil[2]{\small \textit{Department of the Geophysical Sciences, University of Chicago, Chicago, Illinois}}
\affil[3]{\small \textit{Department of Climate, Meteorology, and Atmospheric Sciences, University of Illinois Urbana-Champaign, Urbana, Illinois}}
\affil[4]{\small \textit{Department of Earth Sciences and Environmental Change, University of Illinois Urbana-Champaign, Urbana, Illinois}}
\affil[5]{\small \textit{National Center for Supercomputing Applications, Urbana, Illinois}}
\date{\textit{Preprint -- Comments Welcome}\\\today}

\begin{document}
\maketitle

\begin{abstract}
Climate sensitivity has remained stubbornly uncertain since the Charney Report was published some 45 years ago.
Two factors in future climate projections could alter this dilemma: (i) an increased ratio of CO$_2$ forcing relative to aerosol cooling, owing to both continued accumulation of CO$_2$ and declining aerosol emissions, and (ii) a warming world, whereby CO$_2$-induced warming becomes more pronounced relative to climate variability.
Here, we develop a novel modeling approach to explore the rates of learning about equilibrium climate sensitivity and the transient climate response (TCR) and identify the physical drivers underpinning these learning rates.
Our approach has the advantage over past work by accounting for the full spectrum of parameter uncertainties and covariances, while also taking into account serially correlated internal climate variability. Moreover, we provide a physical explanation of how quickly we may hope to learn about climate sensitivity. 
We find that, although we are able to constrain future TCR regardless of the true underlying value, constraining ECS is more difficult, with low values of ECS being more easily ascertained than high values.
This asymmetry can be explained by most of the warming this century being attributable to the fast climate mode, which is more useful for constraining TCR than it is for ECS.
We further show that our inability to constrain the deep ocean response is what limits our ability to learn high values of ECS.
\end{abstract}

\doublespacing

\section{Introduction}
%Climate sensitivity is an important and policy-relevant characteristic of the climate system.
Broadly speaking, climate sensitivity relates the amount of atmospheric carbon dioxide (CO$_2$) to the amount of global average warming that will result on various timescales. Climate sensitivity can be more precisely defined depending on the timescale under consideration.
On long timescales (e.g., centuries or more), one often considers the equilibrium climate sensitivity (ECS), which quantifies the equilibrium amount of warming that would result from a doubling of atmospheric CO$_2$ relative to pre-industrial times.
On near-term timescales (e.g., years to decades), the transient climate response (TCR) is more common, which quantifies the warming at the point where CO$_2$ has doubled in a simulation where CO$_2$ is increased by 1\% each year and the physical climate has not yet reached equilibrium~\citep{ar6_wg1}.

Though each of these measures of the climate response have qualitative differences, they share a common characteristic of being uncertain owing to the stochasticity and complexity of the climate system.
At the most basic level, the uncertain nature of climate sensitivity, regardless of its precise definition, is the foundation of climate risk, making climate sensitivity a metric of substantive importance in climate policy~\citep{broecker_unpleasant_1987, barnett_pricing_2020, bauer_carbon_2024}.
It immediately follows that any reduction in uncertainty in each of these sensitivity measures will result improved decision-making and, hopefully, cost-savings for policymakers formulating carbon dioxide abatement policy~\citep{hope_policy_1993}.

Frustratingly, narrowing uncertainty in climate sensitivity, regardless of the timescale, has remained a major challenge, with the ``likely" range for ECS not narrowing significantly since the Charney Report published in the 1970s~\citep{national_research_council_carbon_1979, roe_why_2007, sherwood_assessment_2020, ar6_wg1}.
One reason for this is the signal of climate change -- that is, CO$_2$-induced warming relative to preindustrial -- has not been present for long enough in the observational record to place a strong constraint on ECS~\citep{proistosescu_slow_2017}.
Another is that this signal is confounded by a number of factors; the main factor is aerosols that have a very uncertain cooling effect on the global climate, offsetting some amount of CO$_2$ warming and making the task of ascertaining how much of the current climate signal is attributable to CO$_2$ difficult~\citep{knutti_constraints_2002}.

The utility of the observational record in constraining the distribution of climate sensitivity has thus far been limited.
In its sixth assessment report, the Intergovernmental Panel on Climate Change (IPCC) took a multi-pronged approach to constraining climate sensitivity, combining observations, coupled climate models, paleoclimate records, simplified climate emulators, and emergent constraints in a Bayesian framework~\citep{ar6_wg1}. 
Within the context of this estimation framework, the observationally-inferred historical energy budget provides a strong constraint on the low end of ECS, but did not meaningfully constrain the high end of the ECS distribution~\citep{sherwood_assessment_2020}.
This is because of both the state-dependence of climate feedbacks on the pattern of global warming~\citep{zhou_analyzing_2017, dong_attributing_2019} and the confounding effects of aerosol emissions.
The historical observational record can be used to indirectly constrain the high end of the ECS distribution via climate model weighting schemes that down-weight climate models that are `too hot' in the historical record and generally have unrealistically high ECS values~\citep{knutti_end_2010, liang_climate_2020, tokarska_past_2020, ribes_making_2021, hausfather_climate_2022, mcdonnell_what_2024}, but these approaches usually only constrain the extreme tail of the ECS distribution. 

However, two socio-economic trends could break the current stalemate in narrowing climate sensitivity uncertainty using observations.
Each of these trends would increase the CO$_2$-induced warming signal-to-noise ratio, making climate sensitivity, in principle, easier to ascertain.
First, a widespread trend of increasingly stringent air quality laws have caused steady declines in aerosol emissions over the last several decades~\citep{us_environmental_protection_agency_office_of_air_and_radiation_benefits_2011}; indeed, almost all of the shared socio-economic pathways (SSPs) project declining aerosol emissions into the future~\citep{riahi_shared_2017}.
This implies that, if future air quality laws become more stringent, less CO$_2$-induced warming will be masked by the cooling effect of aerosols, thus increasing the signal-to-noise ratio of CO$_2$-attributable warming.
The second factor is that as more CO$_2$ is emitted, the signal of CO$_2$ warming will become more pronounced relative to internal climate variability, making the signal of climate change more easily discernible and bolstering our ability to constrain climate sensitivity using new observations.
%If the relative proportion of CO$_2$ emissions to aerosol emissions continues to rise, this implies that even in the absence of new air quality laws or new technologies, more and more climate signal will be attributable to CO$_2$, making climate sensitivity more easily ascertained owing to a heightened signal-to-noise ratio of CO$_2$-induced warming relative to internal variability.

Another important consideration relates to which climate sensitivity parameter one hopes to learn.
On the one hand, because the characteristic timescales of the climate system's thermal response can exceed two hundred years~\citep{geoffroy_transient_2013-1, leach_fairv200_2021}, near-term warming information is likely to be of limited use in improving our estimates of ECS (as the ``slow" mode of warming will not be manifest for perhaps hundreds of years).
On the other hand, improving our estimates of TCR may yet be possible, as the near-term scope of TCR makes near-term warming information potentially useful, especially in light of the two projected socio-economic trends mentioned above.
This important dynamic trade-off between when the short- and long-term timescales of warming manifest was explored by~\citet{roe_why_2007} and~\citet{baker_shape_2009}. 

Past studies have used synthetic future observations of the climate, either generated with an energy balance model (EBM) or general circulation models (GCM), to constrain estimates of climate sensitivity using Bayesian frameworks~\citep{kelly_bayesian_1999, leach_climate_2007, padilla_probabilistic_2011, j_ring_bayesian_2012, urban_historical_2014, kelly_learning_2015, mori_value_2018}.
Although the specifics of each framework in the literature varies, a common thread is to jointly estimate a subset of the EBM parameter space (e.g., joint uncertainty in the climate feedback, ocean diffusivity, and the sensitivity of aerosol forcing, as done in~\citet{j_ring_bayesian_2012}) to estimate a ``learning rate" of climate sensitivity once future observations are incorporated into the assimilation window.
Most studies find that learning about climate sensitivity is possible by the end of the century, although some suggest that uncertainty may widen before it begins to narrow~\citep{hannart_disconcerting_2013}. 
Another study focused on projecting the decline in anthropogenic radiative forcing uncertainty to constraint the TCR uncertainty, finding that TCR uncertainty could be reduced by as much as 50\% in the near future~\citep{myhre_declining_2015}.
Finally, the Allen, Stott and Kettleborough (ASK) method~\citep{allen_quantifying_2000, stott_origins_2002, gillett_improved_2012, stott_upper_2013} has been used to constrain future temperature rise conditional on sequentially assimilated observations, but a recent application did not explicitly tie narrowing uncertainty in temperature rise to learning about the underlying distribution of ECS or TCR~\citep{shiogama_predicting_2016}.

While the above studies have shed light on the rate at which we narrow uncertainty in climate sensitivity (which we coin ``climate sensitivity learning rates" throughout), they share a number of limitations.
First, current state-of-the-art simplified EBMs (such as FaIR~\citep{leach_fairv200_2021} or MAGICC~\citep{meinshausen_emulating_2011} that are used in the IPCC's estimation methodology mentioned above) are calibrated by treating every parameter in the EBM as uncertain, not just a few that are deemed to be the most vital~\citep{cummins_optimal_2020}.
This implies that different methodologies are being used by those assessing climate sensitivity using EBMs based on past observations~\citep[see, e.g., Chapter 4 in][]{ar6_wg1} and those making predictions about how climate uncertainty will evolve in time.
As a concrete example, we are not currently aware of any study on the learning rate of climate sensitivity that treats the deep ocean heat capacity as uncertain, despite the known difficulty in constraining it~\citep[see, for example,][]{geoffroy_transient_2013-1}. By neglecting certain EBM parameters in the parameter estimation problem, past work may have considerably underestimated the underlying covariance structure between many EBM parameters, which could significantly change climate learning rate estimates both quantitatively and qualitatively.
% MAYBE A SENTENCE ON: IF YOU ONLY TREAT LAMBDA AS UNCERTAIN, WHAT DOES THAT MEAN FOR TAU? Moreover, by only sampling a limited number of parameters in the EBM, past work may be misrepresenting the uncertainty in other aspects of the climate that are important to consider, such as

Second, all previous studies treat the sensitivity of radiative forcing to CO$_2$ concentrations as given, despite known uncertainty in the magnitude of the response of forcing to an injection of CO$_2$ into the atmosphere~\citep{zelinka_causes_2020, ar6_wg1, he_state_2023}.
Past work therefore assumes all uncertainty in ECS comes from uncertainty in the climate feedbacks, despite radiative forcing sensitivity uncertainty strongly trading-off with climate feedback uncertainty in estimates of ECS uncertainty.
Capturing this trade-off would likely alter the qualitative features of climate learning rates.

Lastly, though some studies have pointed to the different characteristic timescales of the climate response to warming as an important factor for climate learning rates~\citep{urban_historical_2014}, explicitly decomposing the learning rates of the different climate modes, and connecting them to the overall climate sensitivity learning rate, remains lacking.
This analysis would provide physical insights into why and how we may expect to learn about the climate in the future rather than relying on computational models alone.

In light of the limitations and challenges described above, this paper presents a modeling framework that merges an IPCC-consistent climate emulator with ensemble variational data assimilation to assess the prospects and perils of learning ECS and TCR from hypothetical future observations of global average temperature and ocean heat content.
Our framework utilizes a weak-constraint, ensemble of data assimilations (EDA) approach similar to those used in the weather forecasting literature~\citep[see, e.g.,][]{isaksen_ensemble_2010} to incorporate ``pseudo-observations" of the future climate to estimate the posterior distribution of climate sensitivity and other climate parameters conditional on future observations (see \textit{Materials and Methods}).

Our modeling framework has two distinct advantages over past work.
The first is we are able to jointly estimate the conditional probability distribution of each parameter in our EBM, as opposed to a subset of our parameter space.
This allows us to capture trade-offs between CO$_2$ radiative forcing sensitivity and climate feedbacks on the posterior distribution of ECS, while also accounting for the impact of difficult-to-learn parameters, such as the deep ocean heat capacity, on our estimates of learning rates.
As we will see later, this has an especially substantive impact on the rate we are able to learn ECS from future observations.
The second is that, by using a so-called ``weak-constraint" variational framework~\citep[WC-VAR, see][and \textit{Materials and Methods}]{evensen_data_2022}, we are able to account for the trade-off between estimating the climate \textit{state} -- that is, what the true climatic temperature is in the absence of internal climate variability -- and estimating the internal climate variability itself~\citep{nicklas_efficient_2024}, a novelty in the climate learning rates literature.
 
\section{Results} \label{sec:results}

\begin{figure}
    \centering
    \includegraphics[width=\linewidth]{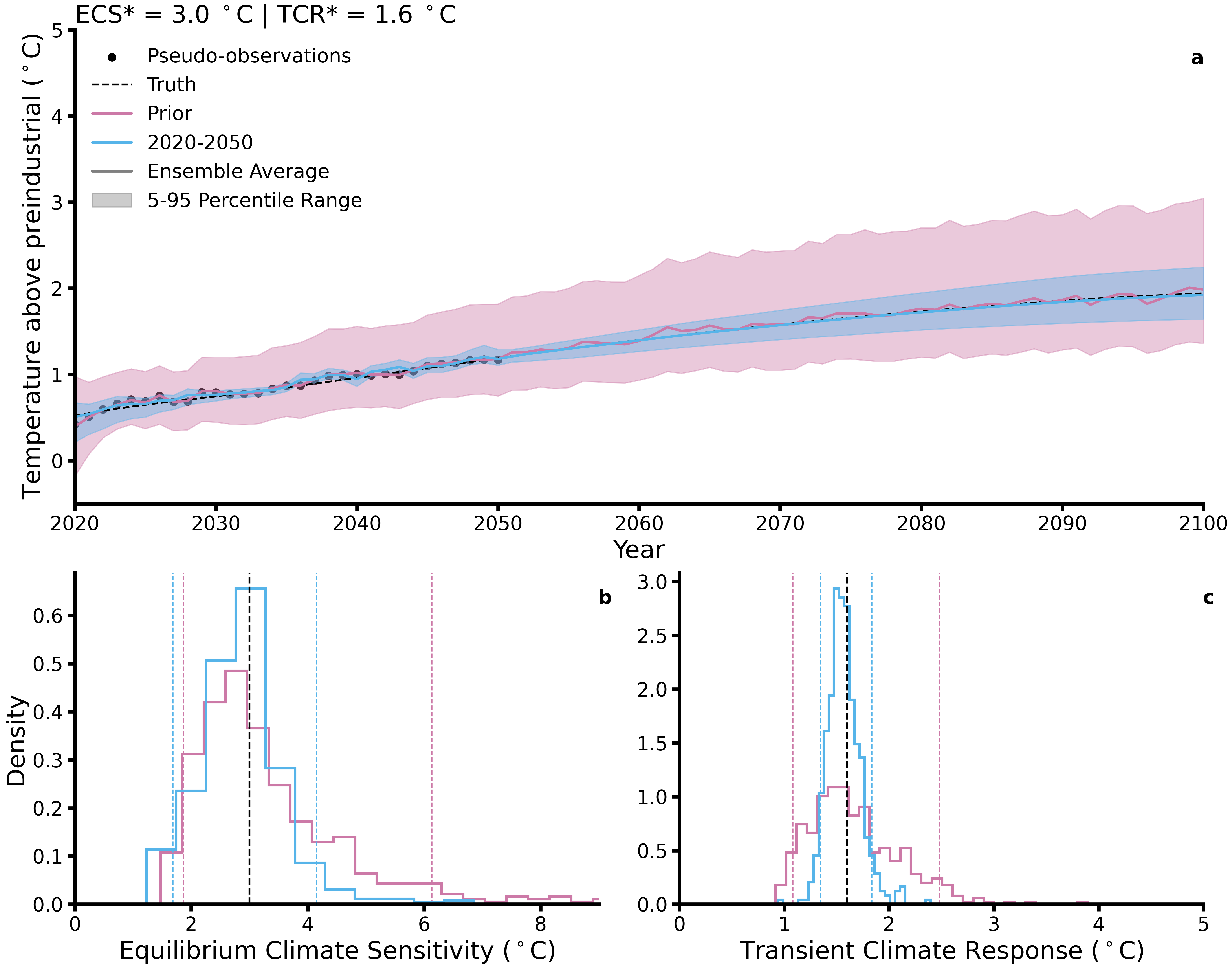}
    \caption{\textbf{Temperature forecasts, ECS, and TCR posteriors conditional on observations up to 2050.} \textbf{a} shows the forecast of future temperatures using parameter posteriors from the 2020-2050 assimilation window (blue) and using the prior distribution (pink) where the true values of ECS and TCR are $3 \ ^\circ$C and 1.6 \textdegree C, respectively. The shaded regions are the 5-95 percentile range, and future temperature pseudo-observations are shown in the black dots. \textbf{b} shows the prior ECS histogram (pink) and the 2020-2050 assimilation window ECS posterior (blue), with the black vertical line representing the true ECS value. \textbf{c} is as \textbf{b}, but for TCR. Pseudo-observations are generated using SSP2--4.5, and in \textbf{b--c}, dashed lines represent the 5-95 percentile range.}
    \label{fig:projections}
\end{figure}

We begin by demonstrating how the forecasted global surface temperature and the distributions of ECS and TCR change as we add observations of global temperature and ocean heat content into our assimilation window (see \textit{Materials and Methods}). We generate pseudo-observations for 2020-2050 by forcing our two-layer model~\citep[][and \textit{Materials and Methods}]{geoffroy_transient_2013-1} with CO$_2$ concentrations and sulfur dioxide (SO$_2$) emissions from SSP2--4.5~\citep{riahi_shared_2017}. We compute the effective radiative forcing using the parameterizations in~\citet{leach_fairv200_2021}. We then treat the time period of 2020-2050 as our ``analysis" period~\citep{park_principles_2022}, over which we minimize our WC-VAR cost function for each of our 500 ensemble members. Because each ensemble member cost function is independent, this procedure results in randomized maximum likelihood (RML) sampling of the posterior conditional probability distribution function (PDF)~\citep{evensen_data_2022}. We then use the posterior distribution to forecast future temperature change.

In Figure~\ref{fig:projections}, we show the projected future temperature rise using the prior distribution (pink), and the forecasted future temperature rise using the posterior parameter distributions from the 2020-2050 assimilation window (blue) for a true values of ECS of 3 $^\circ$C and a true TCR of 1.6 \textdegree C (Fig.~\ref{fig:projections}\textbf{a}). Note these values are equal to our ensemble prior central estimates; we will probe the dependence of our results on what the true underlying value is later. We also compute the posterior distributions of ECS and TCR for the prior distribution and the parameter posterior (Figs.~\ref{fig:projections}\textbf{b--c}). 

We find the estimated range of 2100 warming levels is significantly reduced after observations of 2020-2050 temperature and ocean heat content are assimilated.
Using our parameter priors, the end-of-century warming 5-95 percentile range is 1.6 $^\circ$C.
After incorporating thirty additional years of observations into our assimilation window, this range reduces to about 0.5 $^\circ$C (a 66\% reduction relative to the prior) when the true value of ECS is 3 $^\circ$C.
This suggests that future observations may be able to place a relatively tight constraint on end-of-century temperature rise, consistent with past work~\citep{shiogama_predicting_2016}.

Our ability to constrain end-of-century temperature rise may seem natural given that, in 2050, our proximity to the end of the century is increased, and so the range of possible outcomes has less time to ``spread out".
However, our results would suggest that the narrowing of the end-of-century temperature rise 5-95 percentile range is also in large part due to an improved ability to estimate model parameters which, in turn, constrain ECS and TCR (see Figs.~\ref{fig:projections}\textbf{b--c} and corner plots in the \textit{Supplementary Information}).
We find that by mid-century, our approach provides a fairly robust constraint on the central value of both ECS and TCR.
In particular, we estimate the ECS ensemble median to be 2.86 \textdegree C (5\% error) and estimate a TCR ensemble median of 1.56 \textdegree C (3\% error).
Not only are the ECS and TCR distribution medians estimated relatively well, but we also find nontrivial reductions in uncertainty: the 5-95 percentile range the ECS distribution decreases by 42\%, while the 5-95 percentile range of the TCR distributions decreases by 65\%.
These results -- the narrowing of the uncertainty band for end-of-century temperature rise, ECS, and TCR -- would tentatively suggest that there is some reason for hope in constraining uncertainty in these quantities in the future.

\begin{figure}
    \centering
    \includegraphics[width=\linewidth]{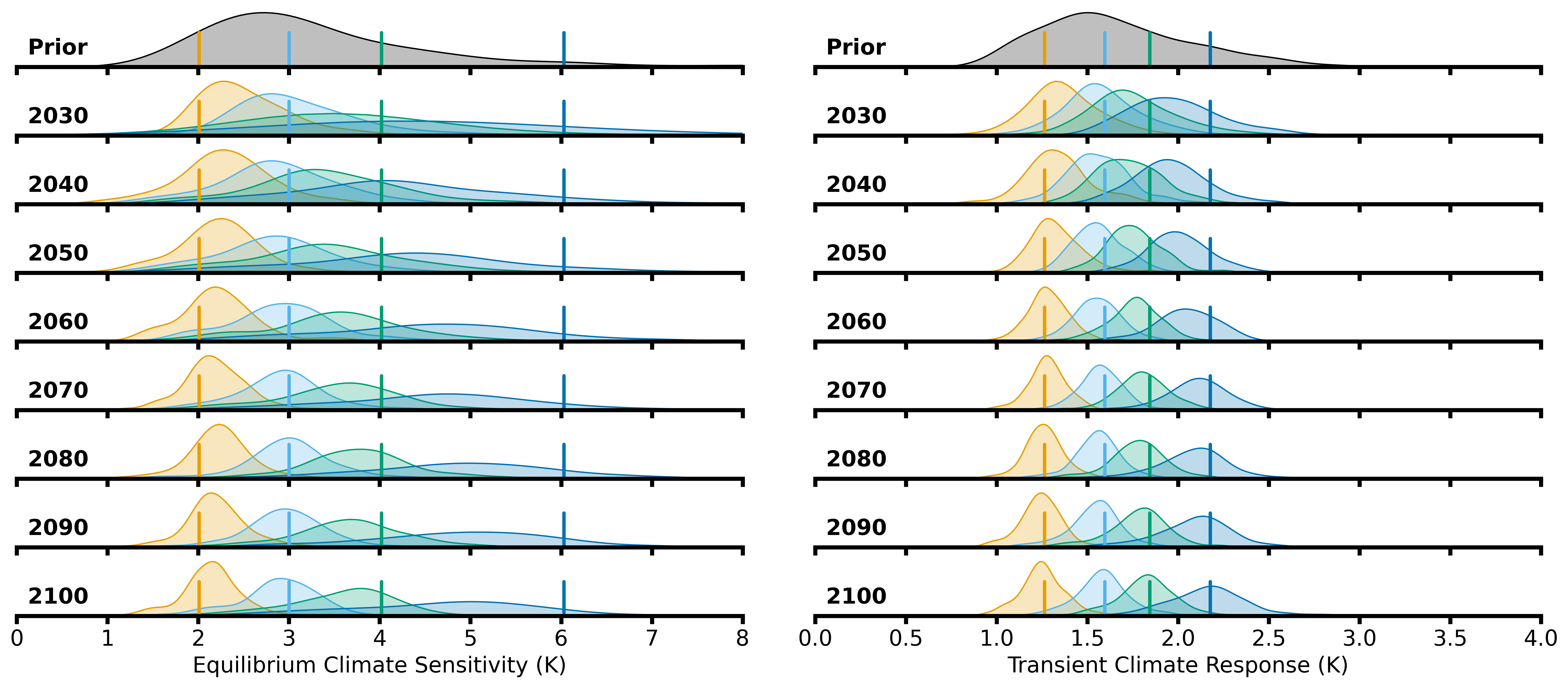}
    \caption{\textbf{Evolution of ECS and TCR posteriors.} On the left, we show how the posterior distribution of ECS evolves for different true values of ECS (solid lines) as the upper bound of our assimilation window is pushed further into the future (and more observations are added to our assimilation window). The right is as the left, but for TCR.}
    \label{fig:ridge}
\end{figure}

Are these results robust regardless of whether we live in an extremely high or low ECS or TCR world?
And how do these results change as more (or less) observations are incorporated into the analysis window?
To answer these questions, we carry out the same exercise demonstrated by Figure~\ref{fig:projections} for different true values of ECS and TCR and display the ECS and TCR posteriors in Figure~\ref{fig:ridge} for numerous analysis window upper bounds.
We find that the degree to which one can ascertain the true value of ECS by the end of the century varies significantly with what the true underlying value is.
For low values of ECS (yellow lines and distributions on the left in Figure~\ref{fig:ridge}), we find that the ECS is learned efficiently by the end of the century, with very little difference in the posterior central estimate and the true value (6.8\% error) and a relatively tight 5-95 percentile range (1.16 \textdegree C).
However, as we increase the true value of ECS, our approach grows increasingly unable to learn the true value; for a high value of ECS ($\sim 5$ \textdegree C, navy lines and distributions in Figure~\ref{fig:ridge}), the percent error between the end of century posterior central estimate and the true value is about 16.2\%, while posterior distribution remains highly uncertain (5-95 percentile range of 2.9 \textdegree C).

However, we find that the persistence of uncertainty and bias in posterior central estimates is only present for ECS.
For TCR, even extremely high values are able to be learned by the end-of-century, with uncertainty narrowing at a similar rate for each true values of TCR we consider (see the right side of Figure~\ref{fig:ridge}).
The percent difference between the true value and the end of century posterior central estimate range from 0.19\%-1.45\%, a far smaller range than that of ECS.
This would suggest that while the prospects for ascertaining \textit{equilibrium} climate metrics, such as ECS, may depend on the underlying true value, \textit{transient} climate metrics, like TCR, do not.
Corner plots in the \textit{Supplementary Information} show that high climate sensitivity experiments struggle to disentangle the trade-offs between the climate feedback strength and the CO$_2$ radiative forcing sensitivity (estimating a too-strong forcing and a too-strong feedback), while low climate sensitivity experiments estimate both parameters effectively. 

%Compared to TCR, our ability to constrain ECS is more limited. Firstly, we find that the bias in the central estimate of ECS depends strongly on the true underlying value. When the true value of ECS of 3 $^\circ$C, the posterior central estimate is about 2.93 $^\circ$C (2\% error), but when the true value is 6 \textdegree C, our 2050 central estimate is 4.6 \textdegree C, underestimating ECS by about 23\%. The degree to which uncertainty declines also varies with the true value of ECS; when the true ECS is 3 \textdegree C, we find a 60\% reduction in the 5-95 percentile range, whereas when ECS is 6 \textdegree C, the 5-95 percentile range is decreased by only 26\%. 

% In sum, these results would suggest that our prospects for constraining TCR by the end of 21\textsuperscript{th} century are far more promising than our prospects of constraining ECS owing to different learning rates. We can further visualize the asymmetric learning rates of ECS and TCR by showing the evolution of the ECS and TCR posteriors as we extend our assimilation window out to 2100 for numerous true values of ECS and TCR, see Figure~\ref{fig:ridge}. 

\begin{figure}
    \centering
    \includegraphics[width=\linewidth]{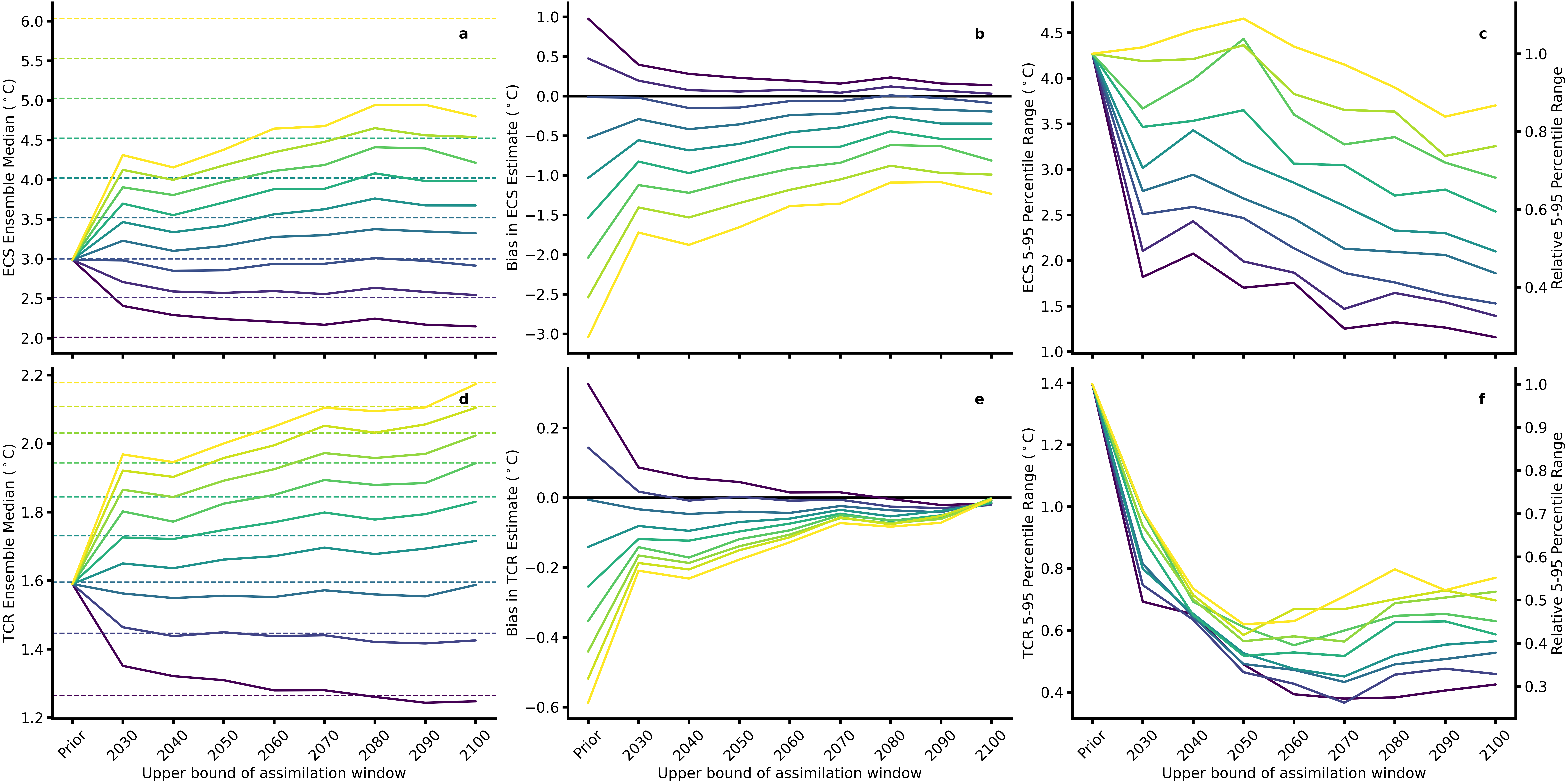}
    \caption{\textbf{ECS and TCR posterior central estimates and learning rates.} \textbf{a} shows the central estimate of the posterior ECS distribution (solid lines) and the true value of the ECS used to generate the observations (dashed lines) as the assimilation window is moved further into the future (and more observations are used for analysis). \textbf{b} shows the difference between the posterior central estimate and the true value, while \textbf{c} shows the raw posterior 5-95 percentile range and the 5-95 percentile range relative to the prior (left and right axes, respectively). \textbf{d--f} are as \textbf{a--c}, but for TCR.}
    \label{fig:central-var}
\end{figure}

We can summarize the above findings by showing the evolution of the posterior central estimate and posterior 5-95 percentile range as more observations are assimilated in Figure~\ref{fig:central-var} for numerous true values of ECS and TCR.
Figure~\ref{fig:central-var} supports the findings of Figure~\ref{fig:ridge}: high values of ECS are unable to be robustly estimated even with new observations through the end of the century (see Figs.~\ref{fig:central-var}\textbf{a, b}), while the bias in the central estimate of TCR becomes virtually negligible by 2100 (Fig.~\ref{fig:central-var}\textbf{e}).

We further find that the ECS 5-95 percentile range decreases at a much slower rate (in relative terms) than the TCR 5-95 percentile range (Figs.~\ref{fig:central-var}\textbf{c, f}).
Indeed, we find that the degree to which uncertainty decreases in the TCR is mostly independent of the true value; for each true value we consider, the 5-95 percentile range is 33\%-44\% that of the prior once observations from 2020-2050 are assimilated, and is roughly constant thereafter.
The ECS posterior, on the other hand, has a 5-95 percentile range between 40\%-109\% of the prior 5-95 range once observations from 2020-2050 are assimilated, and continues to decline somewhat thereafter.
There is also significant stratification between the different true values of ECS; in other words, our ability to narrow uncertainty in ECS is more limited when ECS is high than when it is low.
Hence, not only is our ability to match the ensemble central estimate and the true value different for when we are estimating ECS or TCR, but the rate at which we decrease uncertainty depends on if we are considering an equilibrium or transient climate metric as well.

\begin{figure}
    \centering
    \includegraphics[width=\linewidth]{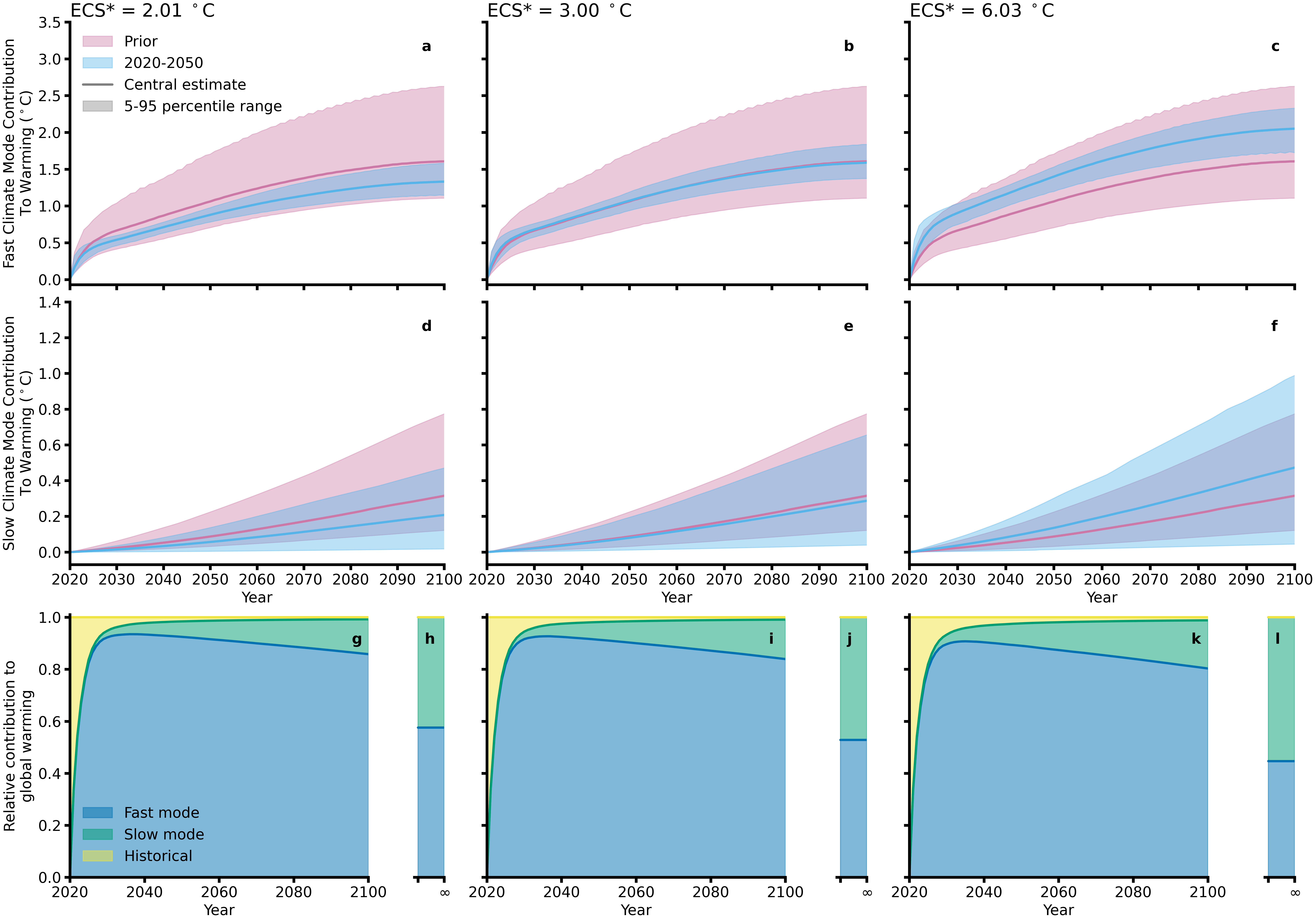}
    \caption{\textbf{Fast and slow mode decomposition of 21\textsuperscript{st} century warming and learning rates.} \textbf{a--c} show the contribution to global warming from the fast climate mode for three different true ECS values (see the titles). \textbf{d--f} is as \textbf{a--c}, but for the slow climate mode. \textbf{g, i, k} show the relative fraction of warming between the fast mode, slow mode, and historical warming for each true value of ECS. \textbf{h, j, l} show the same relative contribution to warming as \textbf{g, i, k}, but after the climate system has reached equilibrium.}
    \label{fig:rel-frac}
\end{figure}

Why can we learn TCR faster and more robustly than ECS, both in terms of the central estimate and the rate of uncertainty decline?
We postulate that the physical intuition behind the differing learning rates displayed in Figure~\ref{fig:central-var} manifests in the relative amounts of warming attributed to the fast and slow climate modes this century.
Using our parameter prior and posterior after assimilating observations from 2020-2050 (\textit{\'a la} Figure~\ref{fig:projections}), we compute the contribution to temperature rise from the fast and slow climate modes in Figure~\ref{fig:rel-frac}\textbf{a--f} for low, average, and extreme values of ECS (see the \textit{Materials and Methods} for details on our decomposition) when forced with SSP2--4.5. Contributions from historical warming are negligible past a decade or two, and have been neglected for our discussion here.
%(see Figure SI.X in the \textit{Supplementary Information}).
We compute the relative fraction of warming from each climate mode in Figure~\ref{fig:rel-frac}\textbf{g--i}. 

We find that most of the uncertainty reduction from thirty years of observations (comparing the pink and blue envelopes in Figure~\ref{fig:rel-frac}\textbf{a--f}) comes from constraining the fast climate mode.
Comparatively, slow mode uncertainty is recalcitrant, experiencing very little reduction in uncertainty (we even find a minor increase in uncertainty in slow mode-contributed warming when ECS is large).
These patterns are consistent regardless of whether or not the true underlying ECS is low or high.
Additionally, the difference in uncertainty reduction between the fast and slow modes happens even while the central estimate of warming attributable to each mode changes to reflect the true climate sensitivity.

The difference in uncertainty reduction for the fast and slow climate mode can be explained by how much near-term warming is attributable to the different modes of warming.
We find that, in the 80 years following the beginning of our simulations, most of the warming found in our pseudo-observations is attributable to the fast climate mode, regardless of the true value of ECS (panels~\ref{fig:rel-frac}\textbf{g--i}).
Physically, this is reflective of the vastly different characteristic timescales of adjustment between the atmosphere and upper mixed ocean layer versus the deep ocean; the mixed ocean layer has an average characteristic timescale of $\sim$10 years, whereas the deep ocean has a characteristic timescale of $\sim$200 years~\citep{geoffroy_transient_2013-1}.
Because warming attributed to the deep ocean manifests over far longer timescales, near-term information on temperature rise is more useful for constraining the fast climate mode, which is the primary contributor to near-term temperature changes and, in turn, the estimated TCR and end of century temperature change~\citep{baker_shape_2009}.

This difference in characteristic timescales between the fast and slow mode of warming can further explain the bias we observed in our central estimates of high values of ECS in Figures~\ref{fig:ridge} and~\ref{fig:central-var}.
We found that if the true underlying value of ECS is high, our central estimate of ECS maintains a lingering bias, even when pseudo-observations are assimilated through the end of the century. This was not the case for low values of ECS. 
The difference between cases where ECS is high or low is related to how much warming is attributable to each climate mode \textit{in equilibrium}, see Figure~\ref{fig:rel-frac}\textbf{g--l}. 
For low values of ECS, the amount of warming attributable to the slow mode in equilibrium is about 42\%, whereas for high values of ECS, the relative amount of warming is about 56\%. This means that for low values of ECS, the deep ocean accounts for less overall warming than the fast mode even after the climate system has equilibrated, whereas the when ECS is high, the deep ocean accounts for the majority of the equilibrium warming. (We prove this explicitly in the \textit{Materials and Methods}.)
Hence, our ability to constrain the fast mode of warming when ECS is low contributes to a tighter constraint on \textit{both} ECS and TCR, while for high values of ECS, the fast mode can only partially constrain ECS while still tightly constraining TCR.

These findings imply that no matter what the true value of ECS is, the information from the next 50 years will be primarily useful for constraining the fast climate mode. In turn, information on the fast climate mode will be primarily useful for constraining transient climate metrics, such as the TCR, and only partially useful for constraining ECS. The remaining information to constrain ECS -- that is, the warming attributable to the slow mode -- will not arrive until hundreds of years hence when the climate system approaches equilibrium, implying that prospects for constraining ECS based on future observations remain limited from a purely physical perspective, and is corroborated by our data assimilation approach.

\section{Discussion} \label{sec:disc}

Our work highlights the key underlying factors important for constraining climate sensitivity from new observations of global temperature and ocean heat content. Our key takeaway is that the prospects for constraining climate sensitivity critically depends on if one is aiming to constrain an equilibrium or a transient climate metric. 

For equilibrium climate sensitivity, we find that our ability to constrain ECS varies depending on the true underlying value; low values of ECS may be constrained by century's end, while high values of ECS are unable to be efficiently estimated even after an additional 80 years of pseudo-observations. Likewise, uncertainty in the distribution of ECS will remain wide for high values of ECS, while for low values of ECS uncertainty may be narrowed by mid-century. The core reason for this asymmetry in the learning rate of ECS for different true values lies in two physical characteristics of the climate: the slow ocean response timescale, which limits the utility of near-term information for constraining ECS, and the fact that the slow mode accounts for a large (small, resp.) share of equilibrium warming when ECS is larger (smaller, resp.) than average. In this way, our work would suggest that if the true value of ECS is large, not only will climate damages be more acute (because temperatures are higher than if ECS is low), but climate risk will also be persistent through the end of the century.
%These findings underscore a potential peril in estimating the ECS from observations in the future: if the true underlying value of ECS is high, then the central estimate of our posterior distributions may continue to underestimate the ECS until well beyond the end of century, and uncertainty will remain stubbornly persistent.

On the other hand, our work would suggest that transient climate metrics, such as end-of-century temperature rise and the transient climate response, may be estimated with some confidence by end of century.
We find that the central estimate of TCR has very little bias by the end of the century regardless of the true underlying value, and uncertainty narrows at a relatively rapid rate. 
This can be explained by end-of-century temperature rise and TCR being primarily controlled by the fast climate mode, which can be tightly constrained in the near-term as CO$_2$-induced warming rises and aerosol emissions decline.
One may therefore be optimistic about narrowing TCR uncertainty in the coming years, despite the prospects for constraining ECS being less easy to judge.

One limitation of our analysis is that we treat our climate feedback parameter and the sensitivity of radiative forcing on CO$_2$ concentrations as independent from the underlying climate state.
However, it has been shown that both of these processes are dependent on the underlying pattern of warming (\citet{zhou_analyzing_2017},~\citet{armour_energy_2017}, and~\citet{dong_attributing_2019} explore the spatial dependence of climate feedbacks, while~\citet{he_state_2023} explores the state dependence of radiative forcing).
In general, it is difficult to capture these effects in global, simplified climate models given their spatial dependence, though we do include an efficacy parameter in our energy balance model that modulates the top of atmosphere energy imbalance to approximate them~\citep{held_probing_2010, cummins_optimal_2020, leach_fairv200_2021}. Including state dependent feedbacks would likely lower our learning rates presented here, given that the transient patterns of warming would obfuscate or ability to learn the ``true" underlying climate sensitivity during the adjustment period. We note that our framework is amenable with a dynamic, time-dependent climate feedback or radiative forcing sensitivity, and future work could incorporate these effects to evaluate their impact on climate learning rates. We expect that while the our quantitative results would be impacted by including these effects, our core physical insight (that transient metrics can be learned more effectively than equilibrium metrics) will still hold in the modified framework.

Our findings have direct implications for climate risk assessment and policymaking. At a technical level, our modeling framework puts forth a methodology to update uncertainty as more pseudo-observations are incorporated into climate model projections. One can image merging our framework with a scenario-based approach to climate risk management in the insurance or financial sector, which would allow climate risk can evolve endogenously with the emissions scenario under consideration, enabling for a more holistic assessment of climate risk in decision-making~\citep[see, e.g.,][]{neubersch_operationalizing_2014}.

For policymakers, the rate at which one can learn about the climate system -- often coined ``active learning" -- has been shown to be important for estimates of the social cost of carbon (SCC)~\citep{kelly_bayesian_1999, leach_climate_2007, kelly_learning_2015, lemoine_managing_2017}.
%Frameworks that estimate this ``active learning" component of the SCC have relied on Bayesian updating about climate feedbacks~\citep{kelly_learning_2015}, but our framework would allow for a reassessment of these modeling frameworks with a fully-fledged climate emulator. 
Our results suggest that the degree to which active learning plays a role in policymaking depends on whether one is considering long- or short-run climate policies. Long-run climate policies often rely on ECS as the key source of climate risk~\citep{weitzman_ghg_2012}, though recently some models utilize the transient climate response to emissions to map cumulative CO$_2$ emissions to temperature rise~\citep{allen_warming_2009, campiglio_optimal_2022}. Short-run policies, on the other hand, often rely on carbon budgets as their main source of climate uncertainty~\citep[e.g.,][]{bauer_how_2024}. Given that the remaining carbon budget is a transient climate metric, it is likely that the quantitative implications of active learning will play a larger role for carbon budget-based policies than ECS-based policies.
% An interesting direction for future work would be to quantify the differences in value of learning equilibrium or transient climate metrics in formulating carbon mitigation policies.

Overall, our analysis suggests that while uncertainty will remain persistent despite new information on the climate over the coming years, we can expect some decline in uncertainty as more observations of the CO$_2$-warming-dominated climate are assimilated into climate models.
Our results provide a physical underpinning for why transient climate metrics are learned at a disproportionally fast rate compared to equilibrium climate metrics.
However, our results are meant to be a stylized demonstration of this fact, and many climatic processes likely will obscure learning from happening as quickly as we have demonstrated here.
% For example, the pattern effect induces state-dependent climate feedbacks~\citep{armour_energy_2017}, which would obfuscate estimates of our climate feedback parameters, especially the heat transfer coefficient in our two-box model.
For example, our quantitative results would need to be reassessed if, say, climate change triggers fundamental process-level changes that are not included in our two-layer box model (e.g., accelerated permafrost melt).
Despite of these limitations, our results suggests that there are fundamental physical reasons to expect that we may be able to constrain climate sensitivity from new observations in the future, and that, depending on the climate metric of interest, these constraints may be substantial.

\section{Materials and Methods} \label{sec:methods}
\subsection{Summary of our modeling approach}
We utilize pseudo-observations and an ensemble of data assimilations (EDA) approach to estimate the rate at which climate sensitivity, and other climate metrics, are learned conditional on future observations of the global climate. Using a large ensemble (in our case, 500 members), we minimize an ensemble of cost functions that allows us to sample the posterior distribution of model states, parameters, and errors conditional on observations of the climate. Throughout, we use a two-layer climate emulator~\citep{geoffroy_transient_2013-1} to compute the global surface temperature and ocean heat content, and use the forcing representation of FaIR v2.0.0~\citep{leach_fairv200_2021} to compute the radiative forcing of each of our forcing agents (e.g., aerosols and greenhouse gases). Our model is forced with CO$_2$ concentrations time series and SO$_2$ emissions time series from SSP2--4.5 (and SSP5--8.5 in a sensitivity analysis)~\citep{riahi_shared_2017}.

We apply our modeling approach in a so-called ``perfect model" context~\citep{urban_historical_2014}, where we generate synthetic observations of the past and future using the same model we attempt to fit the data with. The only difference in the ``perfect model" case is that the observations are forced with a known stochastic component, whereas the retrieved model fit is forced with an unknown forcing that we treat as a control variable (as is standard in WC-VAR). We vary the true underlying value of ECS and TCR by altering the climate feedback parameter, $\lambda$, which allows us to probe how the learning rate about climate sensitivity changes for different true values of ECS or TCR.

\subsection{2-layer energy balance model}
We consider a stochastic two-layer energy balance model (EBM) to emulate the global climate~\citep{geoffroy_transient_2013-1, geoffroy_transient_2013}. We force the EBM with an exogenous path of CO$_2$ concentrations and sulfur dioxide (SO$_2$) emissions, each of which are mapped to an effective radiative forcing through the parameterization of~\citet{leach_fairv200_2021}. In general, if the species of molecules whose radiative forcing are specified by concentrations (emissions, resp.) are given by the set $\mathcal{C}$ ($\mathcal{E}$, resp.) we can write the cumulative radiative forcing at a time \textit{t} as
\begin{align}
    \mathcal{F}_t & = \sum_{c \in \mathcal{C}} f_{1}^{(c)} \ln\left( \dfrac{C^{(c)}_t}{C_{0}^{(c)}}\right) + f_{2}^{(c)} \left( C^{(c)}_t - C^{(c)}_0 \right) + f_{3}^{(c)} \left( \sqrt{C^{(c)}_t} - \sqrt{C^{(c)}_0} \right) \nonumber \\
    & + \sum_{e \in \mathcal{E}} f_{1}^{(e)} \ln\left(1 + \dfrac{E^{(e)}_t}{C_{0}^{(e)}} \right) + f_{2}^{(e)} E^{(e)}_t \label{eq:forcing}
\end{align}
The radiative forcing given by~\eqref{eq:forcing} forces the two layer model given by 
\begin{align}
    C_1 \dfrac{dT_{t}^{(1)}}{dt} & = \mathcal{F}_t - \lambda T_{t}^{(1)} + \varepsilon \gamma \left(T_{t}^{(2)} - T_{t}^{(1)}\right) + q_{t} \label{eq:t1-evo} \\
    C_2 \dfrac{dT_{t}^{(2)}}{dt} & = \gamma \left( T_{t}^{(1)} - T_{t}^{(2)} \right) \label{eq:t2-evo}
\end{align}
where layer \textit{i} has a heat capacity $C_i > 0$ with $i \in \{ 1 , 2 \}$, $\lambda > 0$ is the climate feedback, $\gamma > 0$ is the transfer coefficient between layers, $\varepsilon > 0$ is the ocean heat uptake efficacy, and $q_t$ are model errors. When we use~\eqref{eq:t1-evo}-\eqref{eq:t2-evo} to generate noisy synthetic observations, $q_t = \varepsilon_t$, where $\varepsilon_t$ is a stochastic forcing term that represents internal climate variability. We draw vectors of internal variability such that $\vec{\varepsilon} \sim \mathcal{N}(\vec{0}, \boldsymbol{I})$, where $\boldsymbol{I}$ is the temporal covariance matrix of internal climate variability. In general, the time average of $q_t$ represents model bias,
%(which will be relevant in our ``imperfect model" approach)
 and the temporal covariance represents the covariance structure of internal climate variability.

Importantly, from the integrated paths of~\eqref{eq:t1-evo}-\eqref{eq:t2-evo}, we can compute the ocean heat content, $Q_t$, as
\begin{equation}
    Q_t = C_1 T_{t}^{(1)} + C_2 T_{t}^{(2)}. \label{eq:ocean-heat-content-defn}
\end{equation}
The ocean heat content is a commonly assimilated observation in estimating climate parameters owing to its dependence on the deep ocean temperature, $T_t^{(2)}$. We can then write an amended two layer model where the ocean heat content is also integrated, as
\begin{align}
    C_1 \dfrac{dT_{t}^{(1)}}{dt} & = \mathcal{F}_t - \lambda T_{t}^{(1)} + \varepsilon \gamma \left(T_{t}^{(2)} - T_{t}^{(1)}\right) + q_t \label{eq:t1-evo-f} \\
    C_2 \dfrac{dT_{t}^{(2)}}{dt} & = \gamma \left( T_{t}^{(1)} - T_{t}^{(2)} \right) \label{eq:t2-evo-f} \\
    \dfrac{dQ}{dt} & = \mathcal{F}_t - \lambda T_{t}^{(1)} + (\varepsilon - 1) \gamma \left( T^{(2)}_t - T^{(1)}_t \right) + q_t. \label{eq:q-evo-f}
\end{align}
We discretize~\eqref{eq:t1-evo-f}-\eqref{eq:q-evo-f} using a standard forward Eurler scheme.
%Assuming a time discretization $\Delta t > 0$, the system~\eqref{eq:t1-evo-f}-\eqref{eq:q-evo-f} can be discretized as
%\begin{align}
%    T_{t+1}^{(1)} & = \left( 1 - \dfrac{\Delta t (\lambda + \varepsilon \gamma)}{C_1} \right) T_{t}^{(1)} + \dfrac{\Delta t \gamma \varepsilon}{C_{1}} T_{t}^{(2)} + \dfrac{\Delta t}{C_{1}} \left( \mathcal{F}_t + q_t \right), \label{eq:t1-disc} \\
%    T_{t+1}^{(2)} & = \left( 1 - \dfrac{\Delta t \gamma}{C_2} \right) T_{t}^{(2)} + \dfrac{\Delta t \gamma}{C_{2}} T_{t}^{(1)}, \label{eq:t2-disc} \\
%    Q_{t+1} & = \left( C_1 - \Delta t (\lambda + (\varepsilon - 1) \gamma) \right) T_{t}^{(1)} + \left(C_{2} + \Delta t \gamma (\varepsilon - 1) \right) T_{t}^{(2)} + \Delta t \left( \mathcal{F}_t + q_t \right), \label{eq:q-disc}
%\end{align}
%using a standard forward Euler scheme. 

\subsubsection{Eigenmode decomposition}
Following~\citet{geoffroy_transient_2013-1}, we can use eigenmode decomposition to write the generalized analytical solution to~\eqref{eq:t1-evo}-\eqref{eq:t2-evo}. Notice we may rewrite the system as
\begin{equation}
    \dfrac{d\boldsymbol{T}_t}{dt} = \boldsymbol{D} \boldsymbol{T}_t + \boldsymbol{F}_t, \label{eq:gen-t-evos}
\end{equation}
with
\begin{equation}
    \boldsymbol{T}(t) = \begin{pmatrix}
        T^{(1)}(t) \\
        T^{(2)}(t)
    \end{pmatrix}, \label{eq:t-vector}
\end{equation}
\begin{equation}
    \boldsymbol{D} = \begin{pmatrix}
        -(\gamma \varepsilon + \lambda)/C_1 & \gamma \varepsilon / C_1 \\
        \gamma / C_2 & - \gamma / C_2
    \end{pmatrix}, \label{eq:propagator}
\end{equation}
and
\begin{equation}
    \boldsymbol{F}(t) = \begin{pmatrix}
        \mathcal{F}(t) / C_1 \\
        0 
    \end{pmatrix}. \label{eq:forcing-vector}
\end{equation}
Eqn.~\eqref{eq:gen-t-evos} is well-known to admit a two timescale solution~\citep{proistosescu_slow_2017, strogatz_nonlinear_2018}. Using the variation of parameters method, one can show the analytic solution for the upper layer temperature, $T_t^{(1)}$, is given by
\begin{equation}
    T^{(1)}(t) = T_{hist}^{(1)}(t) + \underbrace{\dfrac{\phi_s}{C_1 (\phi_s - \phi_f)} \int_{0}^{t}\mathcal{F}(\zeta) e^{-(t - \zeta)/\tau_f} d\zeta}_{=: T_{fast}^{(1)}(t)} - \underbrace{\dfrac{\phi_f}{C_1 (\phi_s - \phi_f)} \int_{0}^{t}\mathcal{F}(\zeta) e^{-(t - \zeta)/\tau_s} d\zeta}_{=: T_{slow}^{(1)}(t)},
\end{equation}
where $\tau_f$, $\tau_s$, $\phi_s$, and $\phi_f$ can be found using the quantities given in Table~\ref{tab:eigen-params},
\begin{equation}
    T^{(1)}_{hist}(t) = \dfrac{1}{\phi_s - \phi_f} \left( \left[ \phi_s T^{(1)}_0 - T^{(2)}_0 \right] e^{-t/\tau_f} + \left[ T^{(2)}_0 - \phi_f T^{(1)}_0 \right] e^{-t/\tau_s} \right), \label{eq:t-hist}
\end{equation}
and $T^{(1)}_0$ and $T^{(2)}_0$ are the initial conditions of each box~\citep{geoffroy_transient_2013-1}. We plot each of these components individually in Figure~\ref{fig:rel-frac} using our parameter posteriors.

\begin{table}[]
    \centering
    \caption{\textbf{Summary of eigenmode decomposition parameters and relationships.} Abbreviated version of Table 1 from~\citet{geoffroy_transient_2013-1} including an efficacy term, $\varepsilon$.}
    \begin{tabular}{c c}
    \toprule
    \multicolumn{2}{l}{General parameters} \\
    \midrule
       \multicolumn{2}{c}{$b = \dfrac{\lambda + \varepsilon \gamma}{C_1} + \dfrac{\gamma}{C_2}$} \\
       \multicolumn{2}{c}{$b^* = \dfrac{\lambda + \varepsilon \gamma}{C_1} - \dfrac{\gamma}{C_2}$} \\
       \multicolumn{2}{c}{$\delta = b^2 - 4 \dfrac{\lambda \gamma}{C_1 C_2}$} \\
    \midrule
    \multicolumn{2}{l}{Mode parameters} \\
    \midrule
    Fast & Slow \\
    $\phi_f = \dfrac{C_1}{2\gamma} \left( b^* - \sqrt{\delta} \right)$ & $\phi_s = \dfrac{C_1}{2 \gamma} \left( b^* + \sqrt{\delta} \right)$  \\
    $\tau_f = \dfrac{C_1 C_2}{2 \lambda \gamma} \left( b - \sqrt{\delta} \right)$ & $\tau_s = \dfrac{C_1 C_2}{2 \lambda \gamma} \left( b + \sqrt{\delta} \right)$ \\ 
    \bottomrule 
    \end{tabular}
    \label{tab:eigen-params}
\end{table}

\subsubsection{Fast and slow mode contributions to equilibrium warming}
In the main text, we asserted that higher ECS worlds have more warming that is attributable to the slow climate mode in equilibrium. We now make this notion theoretically concrete. Suppose the fast mode contribution to warming can be well-approximated by the transient climate response, given by
\begin{equation}
    \Delta T_{fast} \approx TCR = \dfrac{F_{2\times}}{\lambda + \gamma}. \label{eq:t-fast-approx}
\end{equation}
Then the slow mode contribution can be written as the remaining warming in equilibrium, i.e.,
\begin{align}
    \Delta T_{slow} & \approx ECS - TCR = \dfrac{F_{2\times}}{\lambda} - \dfrac{F_{2\times}}{\lambda + \gamma}\\
    & = \dfrac{\gamma F_{2\times}}{\lambda (\gamma + \lambda)}. \label{eq:t-slow-approx}
\end{align}
Then the ratio of warming attributed to each mode can be written as
\begin{equation}
    \dfrac{\Delta T_{slow}}{\Delta T_{fast}} = \dfrac{\gamma}{\lambda}. \label{eq:ratio-final}
\end{equation}
Therefore, assuming minimal correlation between $\gamma$ and $\lambda$, in worlds with low $\lambda$ (high ECS), the contribution from the slow mode increases nonlinearly, as supported by the calculations in Figure~\ref{fig:rel-frac}.

\subsection{Ensemble of data assimilations}

The EDA has two primary components: a tangent linear model (TLM) and a cost function, which we develop in turn below. Of particular importance is how a TLM, which can easily be translated into an adjoint model (ADJM), allows for efficient computation of the gradient of the cost function~\citep{errico_what_1997}. Throughout we closely follow~\citet{park_principles_2022} (see their Chapter 4), and refer readers to their exposition for additional technical details. 

\subsubsection{Tangent linear model}
In this section, we introduce the general concept of TLMs, and how they can be verified. In general, any dynamical operator (or ``propagator"), $\boldsymbol{M}_t : \mathbb{R}^N \to \mathbb{R}^N$, that maps a vector of inputs $\boldsymbol{X}_t \in \mathbb{R}^N$ at a time $t$ to the derivative of $\boldsymbol{X}$ can be written as
\begin{equation}
    \dfrac{d\boldsymbol{X}_t}{dt} = \boldsymbol{M}_t\left( \boldsymbol{X}_t \right). \label{eq:gen-forward-model}
\end{equation}
Note that $\boldsymbol{X}_t$ is a collection of model states, and if one is interested in estimating model parameters or model errors, these are included in $\boldsymbol{X}_t$ and are assumed to be stationary; i.e., $d\xi/dt = 0$ for an arbitrary model parameter or error $\xi$. For a small perturbation $\delta \boldsymbol{X}_t$, we can Taylor expand~\eqref{eq:gen-forward-model} as
\begin{align}
    \dfrac{d}{dt} \left( \boldsymbol{X}_t + \delta \boldsymbol{X}_t \right) & = \boldsymbol{M}_t\left( \boldsymbol{X}_t  \right) + \dfrac{\partial \boldsymbol{M}_t}{\partial \boldsymbol{X}_t}\bigg|_{\boldsymbol{X}_t} \delta \boldsymbol{X}_t + \mathcal{O}\left( || \delta \boldsymbol{X}_t ||^2 \right) \\
    & \approx \boldsymbol{M}_t\left( \boldsymbol{X}_t  \right) + \boldsymbol{L}_t \delta \boldsymbol{X}_t \label{eq:taylor-expansion-gen}
\end{align}
where 
\begin{equation}
    \boldsymbol{L}_t := \dfrac{\partial \boldsymbol{M}_t}{\partial \boldsymbol{X}_t}\bigg|_{\boldsymbol{X}_t} \label{eq:tlm-defn}
\end{equation}
is the TLM at time \textit{t} and we have neglected higher order terms. In general, the TLM is a square matrix with shape $N \times N$, where $N$ is the number of control variables one has in their VAR problem. As we will see later, the TLM is of practical importance because of its relationship to the ADJM, and via the ADJM, the gradient of the cost function. To verify that the TLM has been coded correctly and is operating as expected, we can introduce an arbitrary parameter $\alpha > 0$ and use~\eqref{eq:taylor-expansion-gen} to define
\begin{equation}
    R_\alpha := \dfrac{||\boldsymbol{M}_t \left( \boldsymbol{X}_t + \alpha \delta \boldsymbol{X}_t \right) - \boldsymbol{M}_t \left( \boldsymbol{X}_t \right)||_2}{\alpha || \boldsymbol{L}_t \delta \boldsymbol{X}_t ||_2} \label{eq:tlm-check}
\end{equation}
where $|| \cdot ||_2$ is the $L_2$-norm. If the TLM is coded correctly, one will find that $\lim_{\alpha \to 0} R_{\alpha} \to 1$.

\subsubsection{Cost function}

\begin{table}[]
    \centering
    \caption{\textbf{Control variables in our EDA approach.}}
    \begin{tabular}{p{2.8in} p{1in} p{1.7in}}
    \toprule
     Parameter name & Symbol & Initial condition, parameter, or model error \\
    \midrule
     Initial state of surface temperature & $T_{0}^{(1)}$ & Initial condition \\
     Initial state of deep ocean temperature & $T_{0}^{(2)}$ & Initial condition \\
     Climate sensitivity & $\lambda$ & Parameter \\
     Heat transfer coefficient & $\gamma$ & Parameter \\
     Efficacy factor & $\varepsilon$ & Parameter \\
     Surface layer heat capacity & $C_1$ & Parameter \\
     Deep layer heat capacity & $C_2$ & Parameter \\
     Logarithmic sensitivity of concentrations-driven radiative forcing & $f_{1}^{(c)}$ & Parameter \\
     Linear sensitivity of concentrations-driven radiative forcing & $f_{2}^{(c)}$ & Parameter \\
     Square-root sensitivity of concentrations-driven radiative forcing & $f_{3}^{(c)}$ & Parameter \\
     Logarithmic sensitivity of aerosol emissions-driven radiative forcing & $f_{1}^{(e)}$ & Parameter \\
     Aerosol forcing shape parameter & $C_{0}^{(e)}$ & Parameter \\
     Linear sensitivity of aerosol emissions-driven radiative forcing & $f_{2}^{(e)}$ & Parameter \\
     Model errors & $q_t$ & Model error \\
    \bottomrule
    \end{tabular}
    \label{tab:4dvar-params}
\end{table}

The cost function, \textit{J}, can be written in terms of two components: $J_{guess}$, which quantifies the cost of control variables far away from their ``first-guess" estimates, and $J_{obs}$, that penalizes model misfit with observations. Both parts of the cost function are assumed to be derived from Gaussian likelihoods~\citep{evensen_data_2022}.

We begin with $J_{guess}$. Assume that each control variable listed in Table~\ref{tab:4dvar-params} is written as $\vec{x}_0$, and that each control variable as a first-guess of $\vec{\chi}$ with covariance of the joint prior distribution given by $\boldsymbol{P}$. Then we can write $J_{guess}$ as
\begin{equation}
    J_{guess} = \dfrac{1}{2} \left( \vec{x}_0 - \vec{\chi} \right)^T \boldsymbol{P}^{-1} \left( \vec{x}_0 - \vec{\chi} \right). \label{eq:cost-prior}
\end{equation}

As for $J_{obs}$, we write the superposition of two cost functions that quantify deviations in the surface temperature observations and ocean heat content. If $\vec{T}^{(obs)}$ is a vector of observations of surface temperature and $\vec{Q}^{(obs)}$ is a vector of observations of ocean heat content, we can write $J_{obs}$ as
\begin{align}
    J_{obs} & = \dfrac{1}{2}  
        \left( \vec{T}^{(1)} - \vec{T}^{(obs)} \right)^T \boldsymbol{\Sigma}_T^{-1} \left( \vec{T}^{(1)} - \vec{T}^{(obs)} \right) \nonumber \\
        & + \dfrac{1}{2} \left( \vec{Q} - \vec{Q}^{(obs)} \right)^T \boldsymbol{\Sigma}_Q^{-1} \left( \vec{Q} - \vec{Q}^{(obs)} \right) \label{eq:cost-obs}
\end{align}
where $\boldsymbol{\Sigma}_T$ ($\boldsymbol{\Sigma}_Q$, resp.) is the error covariance matrix between observations of surface temperature (ocean heat content, resp.).

Combining~\eqref{eq:cost-prior} and~\eqref{eq:cost-obs}, we can write our cost function as
\begin{align}
    J & = J_{guess} + J_{obs} \nonumber \\
    & = \dfrac{1}{2} \left( \vec{x}_0 - \vec{\chi} \right)^T \boldsymbol{P}^{-1} \left( \vec{x}_0 - \vec{\chi} \right) \nonumber \\
    & + \dfrac{1}{2} \left( \vec{T}^{(1)} - \vec{T}^{(obs)} \right)^T \boldsymbol{\Sigma}_T^{-1} \left( \vec{T}^{(1)} - \vec{T}^{(obs)} \right) \nonumber \\
    & + \dfrac{1}{2} \left( \vec{Q} - \vec{Q}^{(obs)} \right)^T \boldsymbol{\Sigma}_Q^{-1} \left( \vec{Q} - \vec{Q}^{(obs)} \right). \label{eq:cost-all}
\end{align}
The goal of VAR is to minimize this cost function; if we are able to find a global minimum of~\eqref{eq:cost-all}, then our EDA approach is guaranteed to be a randomized maximum likelihood sampling (RML) of the posterior probability distribution function (PDF). Finding the global minimum requires the gradient of the cost function, and we develop a formalism for computing the cost function gradient next.

\subsubsection{Connecting the TLM and the cost function gradient}
The final piece of the VAR approach is to compute the gradient of the cost function~\eqref{eq:cost-all}. Throughout, we will define a vector of control variables at a time \textit{t} as $\vec{x}_{t}$, which is simply a vector containing all of the listed quantities in Table~\ref{tab:4dvar-params}. We will work in two parts, focusing on computing the gradient of $J_{guess}$ and $J_{obs}$ separately; for $J_{obs}$, we will follow the chain rule-based approach of~\citet{Talagrand1991TUo} and~\citet{errico_sensitivity_1992} to compute the gradient of the cost function with respect to the initial condition of the $n^{\mathrm{th}}$ component of $\vec{x}_t$ inductively. 

Starting with $J_{guess}$, the gradient is relatively trivial. Let $x_{0}^{(n)}$ be the $n^{\mathrm{th}}$ component of $\vec{x}_0$, suppose there are $N$ control variables, and let $p_{i,j}$ be the $(i, j)^{\mathrm{th}}$ element of $\boldsymbol{P}^{-1}$. Then directly differentiating~\eqref{eq:cost-prior}, we can write
\begin{equation} 
    \dfrac{\partial J_{guess}}{\partial x^{(n)}_{0}} = p_{n,n} \left( x_{0}^{(n)} - \chi^{(n)} \right) + \sum_{j = 0}^{N} p_{n,j} \left( x_{0}^{(j)} - \chi^{(j)} \right) \label{eq:grad-prior}
\end{equation}
where we have exploited the fact that covariance matrices are symmetric by definition (and therefore so are their inverses).

For $J_{obs}$, we derive the gradient with respect to $x_{0}^{(n)}$ inductively. Let $\sigma^{(T)}_{i,j}$ ($\sigma^{(Q)}_{i,j}$, resp.) be the $(i,j)^{\mathrm{th}}$ component of $\boldsymbol{\Sigma}_T^{-1}$ ($\boldsymbol{\Sigma}_Q^{-1}$, resp.) and define
\begin{equation}
    \Delta^{(T)}_{t} := T_{t}^{(1)} - T_t^{obs}, \label{eq:misfit-T}
\end{equation}
as well as
\begin{equation}
    \Delta^{(Q)}_{t} := Q_{t} - Q_t^{obs}. \label{eq:misfit-Q}
\end{equation}

Starting with the final timestep for which we have model states, $T$, we can write the gradient of $J_{obs}$ with respect to $x_T^{(n)}$ as 
\begin{equation}
    \dfrac{\partial J_{obs}}{\partial x_T^{(n)}} = \left( \sigma_{T,T}^{(T)} \Delta^{(T)}_T + \sum_{\substack{t' = 0 \\ t' \neq T}} \sigma_{t',T}^{(T)} \Delta_{t'}^{(T)} \right) \dfrac{\partial \Delta_{T}^{(T)}}{\partial x_{T}^{(n)}} + \left( \sigma_{T,T}^{(Q)} \Delta^{(Q)}_T + \sum_{\substack{t' = 0 \\ t' \neq T}} \sigma_{t',T}^{(Q)} \Delta_{t'}^{(Q)} \right) \dfrac{\partial \Delta^{(Q)}_{T}}{\partial x_{T}^{(n)}}
\end{equation}
or more succinctly,
\begin{equation}
    \dfrac{\partial J_{obs}}{\partial x_T^{(n)}} = \sum_{k \in \{T, Q\}}\left( \sigma_{T,T}^{(k)} \Delta^{(k)}_T + \sum_{\substack{t' = 0 \\ t' \neq T}} \sigma_{t',T}^{(k)} \Delta_{t'}^{(k)} \right) \dfrac{\partial \Delta_{T}^{(k)}}{\partial x_{T}^{(n)}} =: F_T^{(n)} \label{eq:final-timestep-simplified}
\end{equation}
where we have again exploited the symmetric property of covariance matrices.

We can carry out the same process for $t = T - 1$ and show that
\begin{equation}
    \dfrac{\partial J_{obs}}{\partial x_{T-1}^{(n)}} = F_{T-1}^{(n)} + \sum_{a = 0}^{N} \dfrac{\partial x_{T}^{(a)}}{\partial x_{T-1}^{(n)}} F_{T}^{(a)}. \label{eq:Tless1-simplified}
\end{equation}
Comparing our general formulation of the TLM in~\eqref{eq:tlm-defn} and~\eqref{eq:Tless1-simplified}, we can see that
\begin{equation}
    \sum_{a = 0}^{N} \dfrac{\partial x_{T}^{(a)}}{\partial x_{T-1}^{(n)}} F_{T}^{(a)} \equiv \left(\boldsymbol{L}_T\right)^* \vec{F}_T
\end{equation}
where $\left(\boldsymbol{L}_t\right)^*$ is the adjoint of the TLM. In this way, misfits in the gradient at time $t=T$ are propagated backward in time by the ADJM to $t = T-1$. This approach can be carried out inductively to form a dynamical system, where at every time \textit{t} there is a ``forcing" term $\vec{F}_t$ (given by~\eqref{eq:final-timestep-simplified} where $T$ is replaced by $t$) and a propagated term involving the ADJM and all other forcings for $t' > t$. Integrating this dynamical system back to the initial condition (i.e., from $t = T \to t = 0$) results in the gradient of $J_{obs}$, as desired.

We can check the validity of our ADJM and the gradient of the cost function by introducing two quantities. The first tests the validity of the ADJM; using the definition of the adjoint~\citep{park_principles_2022}, we can write
\begin{equation}
    \Lambda_t := \dfrac{\langle \boldsymbol{L} \delta \boldsymbol{X}, \boldsymbol{L} \delta \boldsymbol{X} \rangle}{\langle \delta \boldsymbol{X} , \boldsymbol{L}^* \left( \boldsymbol{L} \delta \boldsymbol{X} \right) \rangle} \label{eq:adjm-check}
\end{equation}
where $\delta \boldsymbol{X}$ is a perturbed state vector. This identity essentially compares the forward propagated path of the TLM and the backward propagated path of the ADJM. If the program is running correctly, for small numbers of timesteps, $\Lambda_t \approx 1$, and $\Lambda_t$ will diverge from one the more timesteps are taken.

The second quantity we can introduce serves to check our computation of the cost function gradient. For an arbitrary parameter $\alpha > 0$ and vector $\boldsymbol{h}$, we can define
\begin{equation}
    \Phi_\alpha := \dfrac{J\left(\boldsymbol{X} + \alpha \boldsymbol{h}\right) - J\left(\boldsymbol{X}\right)}{\alpha \boldsymbol{h}^T \boldsymbol{\nabla}J\left( \boldsymbol{X} \right)}, \label{eq:cost-grad-check}
\end{equation}
using the definition of a Taylor expansion similar to our derivation of~\eqref{eq:tlm-check}. It is common practice to set $\boldsymbol{h} = \boldsymbol{\nabla} J(\boldsymbol{X})$, but~\eqref{eq:cost-grad-check} is obviously general for any $\boldsymbol{h}$. As was the case with~\eqref{eq:tlm-check}, we expect a correctly formulated cost function gradient to have the limiting behavior of $\lim_{\alpha \to 0} \Phi_\alpha \to 1$.

\subsubsection{Verification of TLM, ADJM, and cost function gradient}

\begin{figure}
    \centering
    \includegraphics[width=\linewidth]{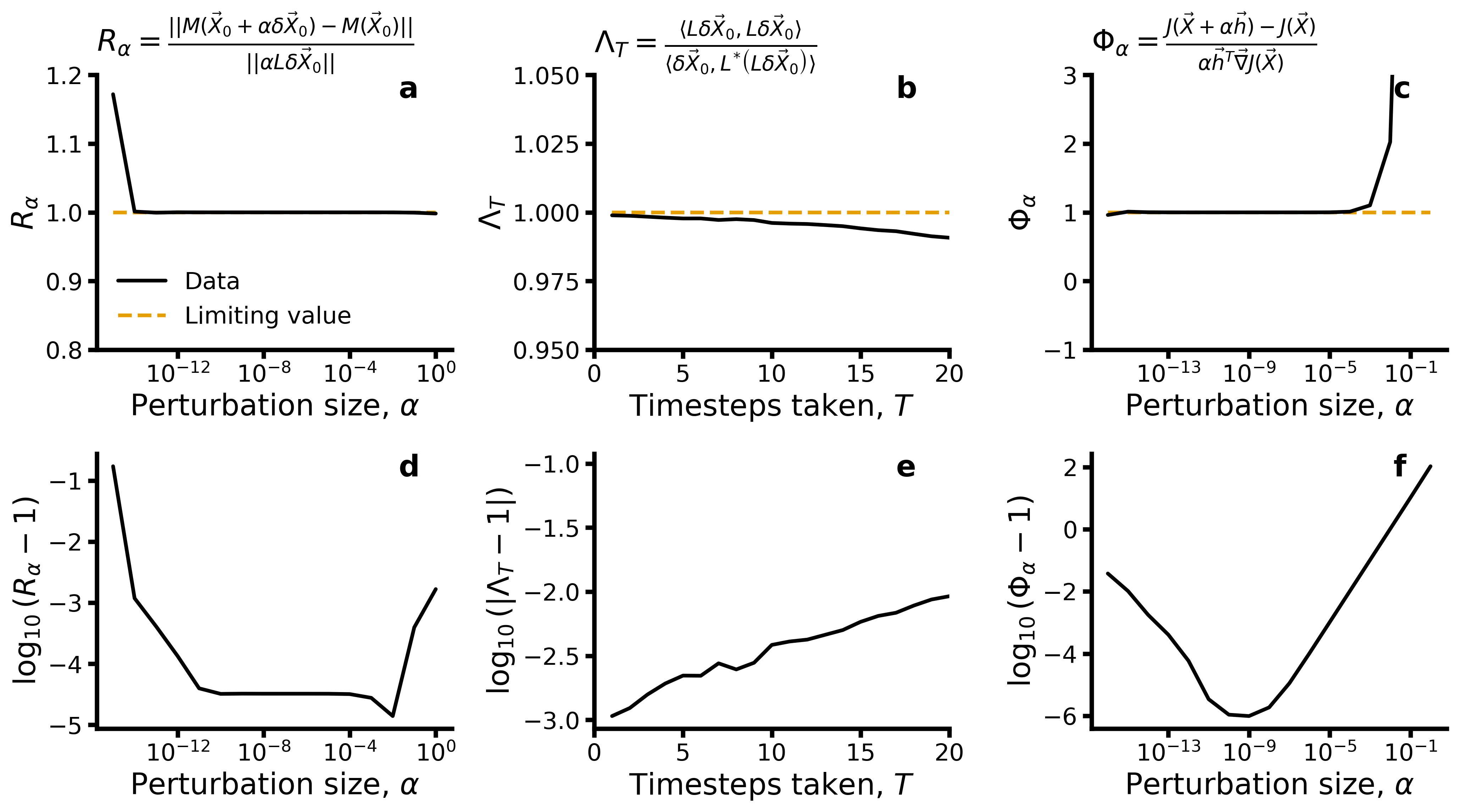}
    \caption{\textbf{Check plots for TLM, ADJM, and cost function gradient.} Panels \textbf{a--c} show~\eqref{eq:tlm-check},~\eqref{eq:adjm-check}, and~\eqref{eq:cost-grad-check} for a number of values of $\alpha$ (or timesteps taken, in the case of panel \textbf{b}). \textbf{d--f} are as \textbf{a--c}, but on a logarithmic scale.}
    \label{fig:check-fig}
\end{figure}

To verify the validity of our formulation, we compute~\eqref{eq:tlm-check},~\eqref{eq:adjm-check}, and~\eqref{eq:cost-grad-check} for a number of perturbation sizes and timesteps in Figure~\ref{fig:check-fig}. We find that each quantity displays the expected behavior for a correctly programmed TLM, ADJM, and cost function gradient. In particular, $R_\alpha$ and $\Phi_\alpha$ approach unity for decreasingly small perturbation sizes, and $\Lambda_t$ starts close to unity for a small number of timesteps before diverging as more are taken. Moreover, on a logarithmic scale, $R_\alpha$ and $\Phi_\alpha$ display the normal ``V"-shaped patterns as is usually found in such computations~\citep{park_principles_2022}.

\subsubsection{Synthesis}
\begin{figure}
    \centering
    \includegraphics[width=\linewidth]{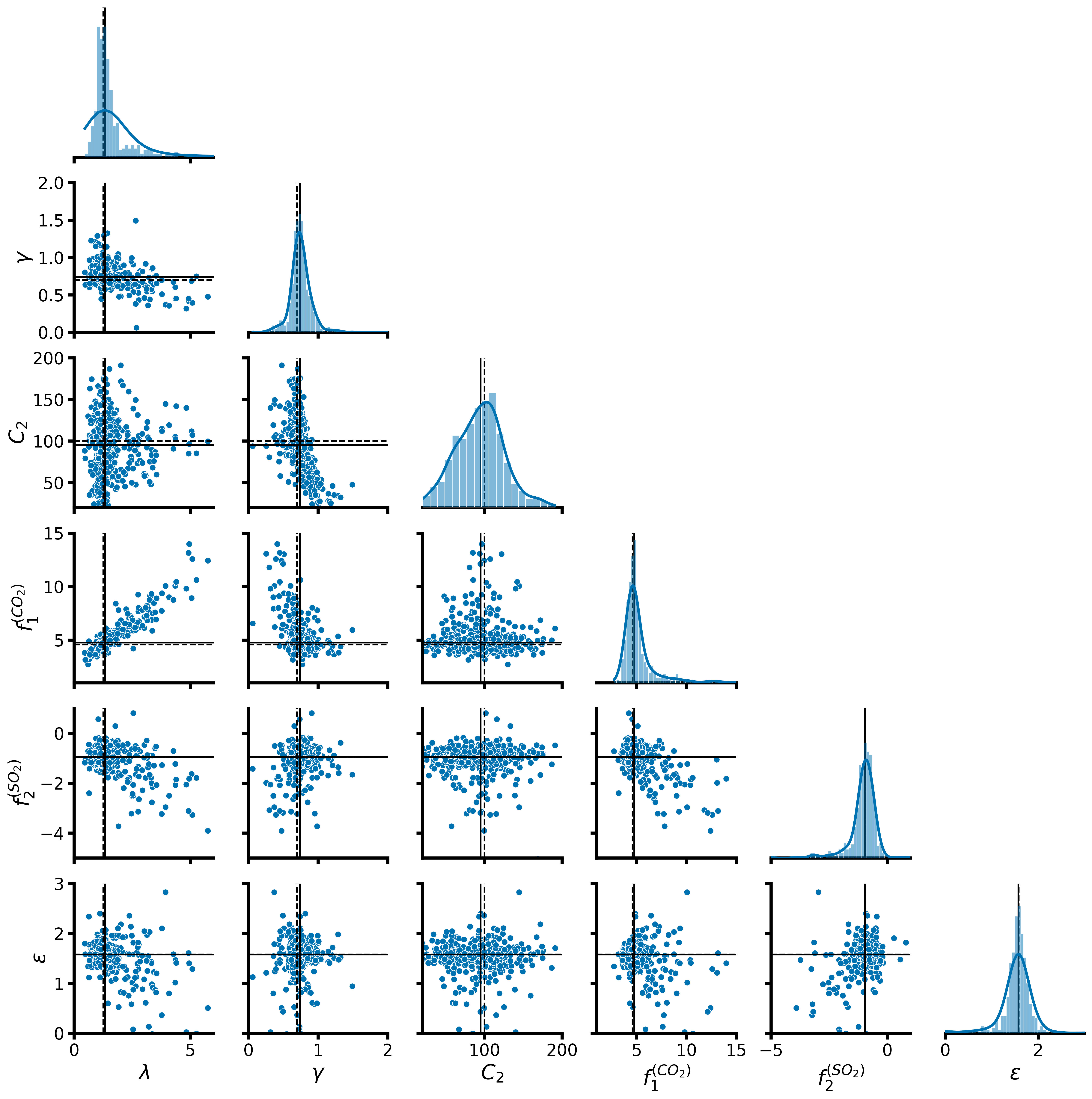}
    \caption{\textbf{Abbreviated parameter posteriors for average true value of ECS.} Black solid lines represent the ensemble median estimate, while the dashed lines represent the true value. Parameter posteriors shown are for the 2020-2060 assimilation window. This is a truncated corner plot where we only show a subset parameters that we estimate in our data assimilation approach: $\lambda$, $\gamma$, $C_2$, $f_1^{(CO_2)}$, $f_2^{(SO_2)}$, $\varepsilon$.}
    \label{fig:corner-abb-avecs}
\end{figure}

We now combine each of the prior subsections to summarize our EDA approach. Assuming the control variables have a prior distribution $\vec{\chi} \sim \mathcal{N}\left(\vec{\chi}^p, \boldsymbol{P}\right)$, we sample this prior $N_{ens}$ times to form an ensemble of ``particles". Each particle's cost function is given by~\eqref{eq:cost-all}, where $\vec{\chi}$ is the prior draw for that particle. This prior draw serves as the first ``first-guess" in the VAR inner- and outer-loop procedure~\citep{park_principles_2022}; it is then updated in the outer-loop of the VAR gradient descent algorithm as the algorithm searches for better solutions to the minimization. We use sequential least-squares programming to solve the inner-loop minimization problem~\citep{noceda_numerical_2006}. An example of an abbreviated corner plot when the true ECS is average is shown in Figure~\ref{fig:corner-abb-avecs} where we show the abbreviated parameter posteriors for the 2020-2050 assimilation window; see the \textit{Supplementary Information} for more corner plots when ECS is low and high, as well as the full parameter corner plots. See~\citet{Rabier2003VariationalDA} and~\citet{bannister_review_2017} for a review of the technical aspects of the VAR approach.

\begin{table}[]
    \centering
    \caption{\textbf{Cost function summary statistics.} Average median, 95\textsuperscript{th} percentile, 97\textsuperscript{th} percentile, 98\textsuperscript{th} percentile, and 99\textsuperscript{th} percentile for each finite posterior cost function distribution, computed for each true value of ECS and assimilation window upper bound we consider.}
    \begin{tabular}{c c}
    \toprule
    Quantity & $\log_{10}\left(\mathrm{Value}\right)$ \\
    \midrule 
    Median & 0.506 \\
    95\textsuperscript{th} percentile & 1.433 \\
    97\textsuperscript{th} percentile & 1.900 \\
    98\textsuperscript{th} percentile & 2.485 \\
    99\textsuperscript{th} percentile & 13.137 \\
    \bottomrule
    \end{tabular}
    \label{tab:cost-func-summary-stats}
\end{table}

\begin{figure}
    \centering
    \includegraphics[width=0.7\linewidth]{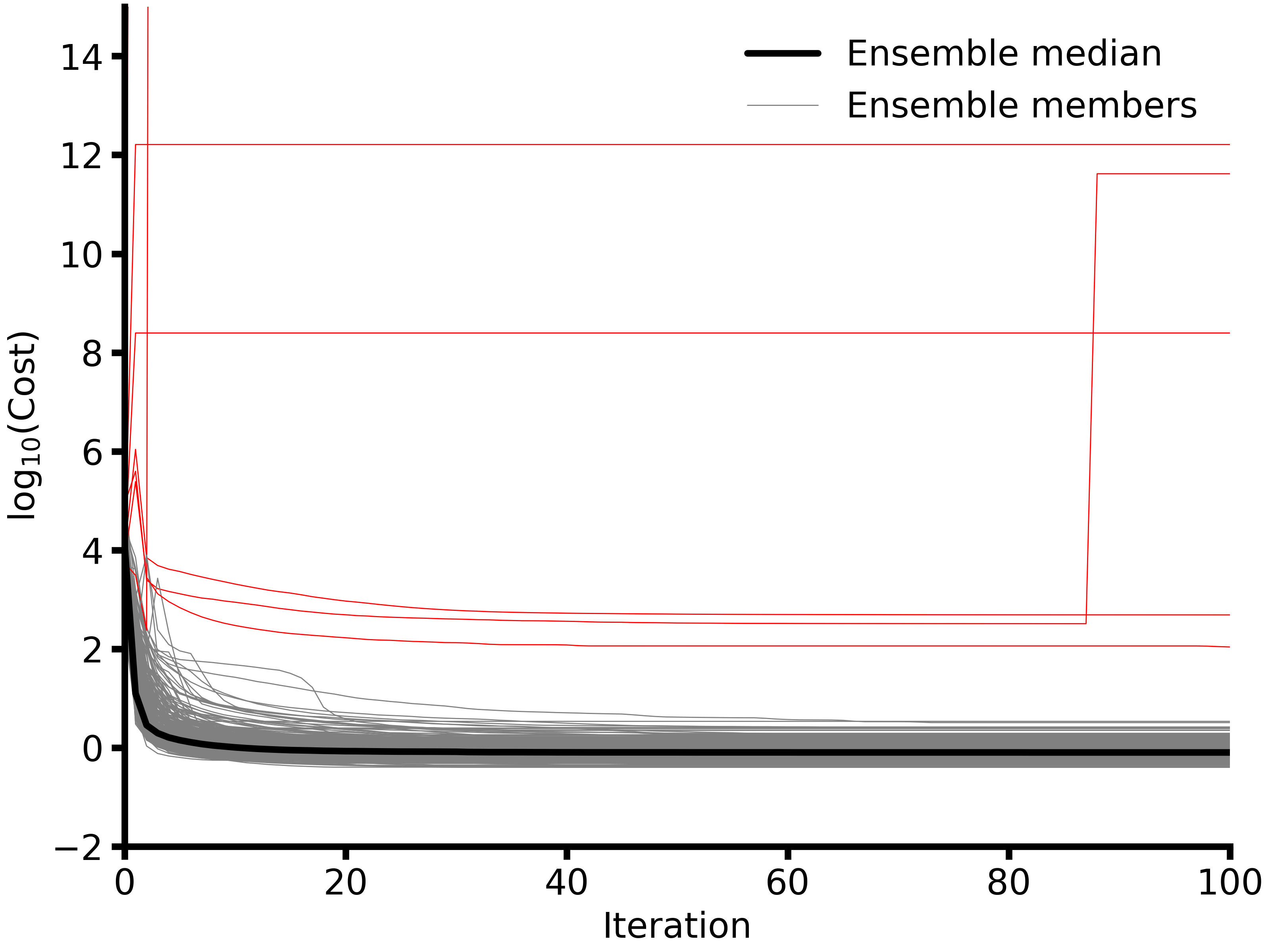}
    \caption{\textbf{Cost function versus iteration.} We plot the cost function for each ensemble member against the VAR iteration number, where valid ensemble members are shown in gray, the ensemble median is in black, and rejected ensemble members are in red. These data are generated for our two-layer model forced with SSP2--4.5 with a true ECS of 6 \textdegree C. Note the upper bound of the assimilation window is 2090.}
    \label{fig:costs}
\end{figure}

We minimize each cost function using VAR, utilizing the cost function gradient derived using~\eqref{eq:grad-prior} and by integrating~\eqref{eq:Tless1-simplified} recursively backwards in time. The result is an optimized ensemble of state estimates that allow us to sample the posterior of our model states, model parameters, and model errors conditional on our data in the assimilation window; indeed, insofar as we find global minima of~\eqref{eq:cost-all}, we are guaranteed a randomized maximum likelihood sampling of the posterior since each particle's cost function is independent~\citep{evensen_data_2022}. We establish a convergence criterion that, after 100 iterations of VAR, the cost function must be below 10\textsuperscript{2}, which places the cut-off just over the average 97\textsuperscript{th} percentile in our summary statistics, see Table~\ref{tab:cost-func-summary-stats}. See Figure~\ref{fig:costs} for an example of our screening methodology.

\subsection{Details on prior}
\begin{sidewaystable}[]
    \centering
    \caption{\textbf{Information on prior calibration.}}
    \begin{tabular}{c c c c}
        \toprule
         Parameter and units & Prior distribution & Source & Notes \\
        \midrule
         $T^{(1)}_0$ [K] & $\mathcal{N}\left(0.46, 0.2\right)$ & Model warm start & -- \\
         $T^{(2)}_0$ [K] & $\mathcal{N}\left(0.06, 0.2\right)$ & Model warm start & -- \\
         $\lambda$ [W K$^{-1}$] & $\mathcal{N}\left(1.258, 0.38\right)$ & \citet{geoffroy_transient_2013-1} & -- \\
         $\gamma$ [W K$^{-1}$] & $\mathcal{N}\left(0.7, 0.21\right)$ & \citet{geoffroy_transient_2013-1} & -- \\
         $C_1$ [J K$^{-1}$] & $\mathcal{N}\left(8, 2.4\right)$ & \citet{geoffroy_transient_2013-1} & -- \\
         $C_2$ [J K$^{-1}$] & $\mathcal{N}\left(100, 30\right)$ & \citet{geoffroy_transient_2013-1} & -- \\
         $f_1^{(\mathrm{CO}_2)}$ [W m$^{-2}$] & $\mathcal{N}\left(4.58, 0.519\right)$ & \citet{leach_fairv200_2021} & Var. from \citet{zelinka_causes_2020} \\
         $f_3^{(\mathrm{CO}_2)}$ [W m$^{-2}$ (ppm CO$_2$)$^{-1/2}$]& $\mathcal{N}\left(0.086, 0.026\right)$ & \citet{leach_fairv200_2021} & -- \\
         $f_1^{(\mathrm{SO}_2)}$ [W] & $\mathcal{N}\left(-0.0047, 0.0014\right)$ & \citet{leach_fairv200_2021} & -- \\
         $f_2^{(\mathrm{SO}_2)}$ [W (MtSO$_2$ / yr)$^{-1}$] & $\mathcal{N}\left(-0.96, 0.29 \right)$ & \citet{leach_fairv200_2021} & -- \\
         $C_0^{(\mathrm{SO}_2)}$ [MtSO$_2$ / yr] & $\mathcal{N}\left(170.6, 51.2\right)$ & \citet{leach_fairv200_2021} & -- \\\
         $\varepsilon$ [--] & $\mathcal{N}(1.58, 0.128)$ & \citet{cummins_optimal_2020} & -- \\
         $q_t$ [W] & $\mathcal{N}(0, 0.27)$ & \citet{proistosescu_slow_2017} & AR(1), $\varphi = 0.2$ \\ 
         \bottomrule
    \end{tabular}
    \label{tab:prior-info}
\end{sidewaystable}

It remains to describe how we chose our prior for each parameter we estimate in our approach. For each box's initial condition, $T^{(1)}_0$ and $T^{(2)}_0$, we warm start our model starting in 1850, and integrate the model until 2020 using the true values of each other parameter in our model. We then assign a 0.2 \textdegree C standard deviation to the prior distribution to account for internal variability and the possibility of errors in measuring the current global average temperature.

We pull the central estimates for our thermal response parameters ($\lambda$, $\gamma$, $C_1$, and $C_2$) from~\citet{geoffroy_transient_2013-1}. Note we slightly revise upwards the estimate of $\lambda$ (1.13 $\to$ 1.258) to achieve an average ECS of 3 \textdegree C, in line with recent estimates~\citep{sherwood_assessment_2020}. However,~\citet{geoffroy_transient_2013-1} calibrates their two-layer model on only 16 climate models, which calls into question their reported standard deviations for each of these parameter distributions, especially given the correlation structure they report between parameters in their posteriors (see their Tables 2 and 3). These factors complicate the prior for the thermal response parameters. We therefore take a more na\"ive approach: to assign a standard deviation of 30\% error for each parameter. This procedure results in an approximately similar distribution for $\lambda$, a wider distribution for $\gamma$ and $C_1$, and a slightly narrower distribution for $C_2$. Sensitivity tests with a 40\% spread in the prior yield similar results as what was reported in the main text (see the \textit{Supplementary Information}).

We pull the central estimate and standard deviation of $f_{1}^{(CO_2)}$ from~\citet{zelinka_causes_2020}, their Table S1.
Note that~\citet{zelinka_causes_2020} reports the central estimate and variance of effective radiative forcing from a doubling of CO$_2$ in CMIP6 climate models; their values were converted into a central estimate and standard deviation for the $f_{1}^{(CO_2)}$ parameter by dividing by $\ln2$.

For the remaining radiative forcing parameters ($f_3^{(CO_2)}$, $f_1^{(\mathrm{SO}_2)}$, $f_2^{(\mathrm{SO}_2)}$, and $C_0^{(SO_2)}$), we pull the central estimates from~\citet{leach_fairv200_2021}.
Unfortunately,~\citet{leach_fairv200_2021} does not report individual parameter uncertainties, and only the uncertainty in overall radiative forcing from different forcing agents.
We therefore take a similar approach to the above and assign a 30\% error for each parameter.
See the below for sensitivity tests; we find that a wider prior results in similar findings.
Note that $f_2^{(CO_2)} = 0$, following~\citet{leach_fairv200_2021}.

Finally, we calibrate the internal climate variability using the AR(1) process following~\citet{proistosescu_slow_2017}. We follow their formulation to write variability in the radiative forcing, $q_t$, as an AR(1) process such that
\begin{equation}
    q_{t+1} = \varphi q_t + \varepsilon_{t+1}
\end{equation}
where $\varepsilon_t \sim \mathcal{N}\left( 0 , \sigma \right)$ and $\varphi \geq 0$. We take the central estimate for $\sigma$ from~\citet{proistosescu_slow_2017}, and perform sensitivity tests for higher value of $\sigma$ (0.4 W m$^{-2}$) below. We take a slightly higher value of $\varphi$ (0.2, compared to 0.03) in our calibration compared to~\citet{proistosescu_slow_2017} to account for the higher autocorrelation in global temperature than top of atmosphere radiative imbalance. See Table~\ref{tab:prior-info} for a summary.

\subsection*{Acknowledgments}
The authors would like to firstly thank Jidong Gao for his help in setting up, verifying, and utilizing our data assimilation approach; this manuscript would not have been possible without his patience and insight. The authors would also like to thank Dan Amrhein, Alfonso Ladino, Joseph Neid, Massimo Pascale, Chris Smith, Ryan Sriver, as well as participants at the American Geophysical Union's 2024 fall meeting for helpful discussions related to this work. AMB acknowledges support from a National Science Foundation Graduate Research Fellowship grant No. DGE 21-46756. Computations were performed on the Keeling computing cluster, a computing resource operated by the School of Earth, Society and Environment (SESE) at the University of Illinois Urbana-Champaign.

\subsection*{Data Availability Statement}
All the code needed to reproduce the analysis of this paper is located on the author's Github, see \texttt{https://www.github.com/adam-bauer-34/BPD-Climate-Learning-Rates-Reprod}.

\subsection*{Supplementary Information}
See the lead author's website, \texttt{https://www.ambauer.com/project/4dvar}.

\bibliographystyle{abbrvnat}
\bibliography{4dvar}

\end{document}